\documentclass[twocolumn,prb,showpacs,color,superscriptaddress,epsfig]{revtex4}
\usepackage{graphicx}
\usepackage{amsfonts}
\usepackage{amsmath}
\usepackage{amssymb}
\usepackage{epsfig}
\usepackage{color}
\definecolor{red}{rgb}{1,0,0}

\begin{document}
\title{Electronic structure and magnetic properties of
Li$_2$ZrCuO$_4$ - a spin 1/2 Heisenberg system in vicinity to a quantum
critical point
}

\author{M. Schmitt}
\affiliation{Max-Planck-Institut f\"ur Chemische Physik fester Stoffe,
  N\"othnitzer Str.\ 40, D-01187 Dresden, Germany}
\author{J. M\'{a}lek}
\affiliation{Institute of Physics, ASCR, Prague, Czech Republic}
\affiliation{Leibniz-Institut f\"{u}r Festk\"{o}rper- und
Werkstoffforschung Dresden, P.O.\ Box 270116, D-01171 Dresden,
Germany}
\author{S.-L. Drechsler}
\affiliation{Leibniz-Institut f\"{u}r Festk\"{o}rper- und
Werkstoffforschung Dresden, P.O.\ Box 270116, D-01171 Dresden,
Germany}
\author{H. Rosner}
\affiliation{Max-Planck-Institut f\"ur Chemische Physik fester Stoffe,
  N\"othnitzer Str.\ 40, D-01187 Dresden, Germany}
%
%
\begin{abstract}
Based on density functional calculations, we present a detailed
theoretical study of the electronic structure and the magnetic
properties of the quasi-one dimensional chain cuprate Li$_2$ZrCuO$_4$ 
(Li$_2$CuZrO$_4$).
For the relevant ratio of the next-nearest neighbor exchange $J_2$ to
the nearest neighbor exchange $J_1$ we find $\alpha = -J_2/J_1 =
0.22\pm0.02$ which is very close to the critical point at 1/4.  Owing
this vicinity to a ferromagnetic-helical critical point, we study
in detail the influence of structural peculiarities such as the reported Li
disorder and the non-planar chain geometry on the magnetic
interactions combining the results of LDA based tight-binding models
with LDA+$U$ derived exchange parameters. Our investigation is
complemented by an exact diagonalization study of a multi-band Hubbard
model for finite clusters predicting a strong temperature dependence
of the optical conductivity for Li$_2$ZrCuO$_4$.
\end{abstract}
\maketitle
\section{Introduction}

Low dimensional magnetism has always attracted great interest in solid
state physics and chemistry. Especially, the phase diagram of low
dimensional spin-1/2 lattices has been investigated extensively, both
theoretically and experimentally. For low dimensions and spin-1/2, the
influence of quantum fluctuations becomes crucial for the ground state
of the system. The role of quantum fluctuations is even more
pronounced if the system under consideration exhibits strongly
frustrated interactions.

Taking only nearest neighbor interactions into account, pure
geometrical frustration in two dimensions (2D) can be realized by
special symmetries, i.e triangular or Kagom\'e lattices. Prominent
real material realizations for such lattices are
$\kappa$-(BEDT-TTF)$_2$X (Ref.\onlinecite{McKenzie}) and
ZnCu$_3$(OH)$_6$Cl$_2$.\cite{Misguich} Another possibility to realize
frustrated couplings are competing nearest neighbor (NN) and next
nearest neighbor (NNN) interactions. This way, the unfrustrated NN
Heisenberg square lattice becomes frustrated by adding
antiferromagnetic (AFM) interactions to NNN. The ground state of this
model is determined by the ratio 
$\alpha$=$-J_2/J_1$ between NNN and NN
exchange interaction. Including a tiny, but non-vanishing coupling
between the magnetic planes that is always present in real materials,
for small $\alpha$ an antiferromagnetic 
ground state is observed like in
all undoped HTSC parent compounds.\cite{Johnston} For large $\alpha$ the so
called columnar order is expected, as found in Li$_2$VSiO$_4$ or
Li$_2$VGeO$_4$.\cite{Melzi00, Melzi01, Rosner02, Rosner_prb03,
  Bombardi} Between these commensurably ordered phases, theory
predicts a spin liquid ground state. The possible realization of this 
spin liquid ground state in PbVO$_3$ is presently under debate.\cite{Shpan04, Oka08}

In one dimension (1D), pure geometrical frustration due to NN exchange
is impossible, but frustration by competing NN and NNN exchanges may
occur in close analogy to the 2D square $J_1$-$J_2$ model. The phase
diagram of this seemingly simple model is very rich. Depending on the
frustration ratio $\alpha$, a variety of ground states was observed in
corresponding quasi-1D systems: (i) ferromagnetically ordered chains
in Li$_2$CuO$_2$ \cite{Nitzsche} ($0\leq \alpha \leq 0.25$, within an
effective single-chain analysis ignoring the antiferromagnetic
inter-chain interaction\cite{footnote1}), (ii) helical order with
different pitch angles in LiVCuO$_4$, LiCu$_2$O$_2$, and NaCu$_2$O$_2$
\cite{Enderle, Masuda, Gippi04, Drechs05, Capogna, Drechs06} ($\alpha
> 0.25$) or spin gap behavior like in the famous spin-Peierls compound
CuGeO$_3$
\cite{Hase} ($-\alpha>0.241$), i.e.~both
exchange couplings are antiferromagnetic.

Very recently, large interest focused on chain systems that are close to the
quantum critical point (QCP) at $\alpha = 0.25$. In the vicinity of a
QCP, the system is expected to answer in a pronounced way to small
external perturbations like magnetic fields or pressure. In addition,
even very small additional exchange paths, especially the inter-chain
coupling, become important for the magnetic properties of the system.
A most promising compound to study the close vicinity to the QCP is
Li$_2$ZrCuO$_4$, where a frustration ratio of $\alpha = 0.29$ was
reported from the evaluation of susceptibility and specific heat
data.\cite{Drechs07} The expected strong field dependence of the
specific heat from magnetic fields up to 9 Tesla was observed,
although the NN exchange $J_1$ was estimated to be rather large in the
order of 300 K. The vicinity of Li$_2$ZrCuO$_4$ to the QCP was
supported by preliminary band structure calculations that obtained a
frustration ratio $\alpha \sim$ 0.23.

In this paper, we report a detailed electronic structure study based
on density functional calculations. In particular, we investigate the
influence of the experimentally observed Li disorder \cite{Duss02} for
one of the crystallographic Li sites on the magnetic properties. We
discuss the dependency of the exchange integrals and the
frustration ratio $\alpha$ 
on the strong Coulomb repulsion within
the Cu $3d$ orbitals. In addition, we elucidate the crucial importance
of the structural distortion of the CuO$_2$ chains in
Li$_2$ZrCuO$_4$. 
This distortion is responsible for a reduction of both the
inter-chain coupling and the NNN exchange $J_2$, 
as a consequence placing the system very close to the QCP.

\section{Methods}

For the electronic structure calculation a full-potential
non-orthogonal local-orbital minimum-basis scheme 
(FPLO5.00-19 and FPLO7.00-28),
\cite{footnote2,Koep1} within the local (spin) density approximation (L(S)DA)
was used. In the scalar relativistic calculations 
the exchange and
correlation potential of Perdew and Wang was used.\cite{PerdW} 
We also applied the general gradient approximation 
(Perdew-Burke-Ernzerhof\cite{Burke96}) for the exchange and correlation 
potential to check whether this influences  
the LDA results. Neither for the band structure nor for the total 
energy differences we found any significant changes.
For
the basis the following states were taken into account: Cu $(3s 3p) /
4s 4p 3d $, Li $ 1s / (2s 2p) + 3d$, O $(2s 2p 3d)$, Zr $(4s 4p) / (5s
5p 4d) $ (notation: \emph{semi-core states / valence states}). The Cu
$3s3p$, Li $1s$ and Zr $4s4p$ states were treated as semi-core states
due to the large extension of their wave functions. The unoccupied
states Li $2p3d$, O $3d$ and Zr $5p$ were considered as hybridization
states to complete the basis. All lower laying states were treated as
core states. To ensure accuracy of the total energy 
300 k-points in the irreducible part
of the Brillouin zone were used.

To model the Li(1) split position of the system (see below), we
employed the virtual crystal approximation (VCA) and coherent potential
approximation (CPA). Alternatively super-cells with different Li(1)
orders have been studied.  To treat the strong on-site Coulomb
repulsion for Cu 3$d$ orbitals explicitly, the 
\mbox{L(S)DA + $U$}
method \cite{Esch1} with $U_{3d}$ in the representative range from
5.5\,eV to 8.0\,eV and $I$ = 1\,eV for the intra atomic exchange have been used.

To study some aspects of 
the magnetic properties of the system a tight binding model
(TBM) was derived from the LDA-calculations
to find the transfer integrals for a single-band effective 
extended Hubbard model 
which subsequently has been mapped onto a
corresponding 3D Heisenberg model.

In addition, exact diagonalization studies of 
a five-band $pd$ extended 
Hubbard model and its mapping onto the 1D-Heisenberg model 
using finite open Cu$_n$O$_{2n+2}$  
clusters have been carried out.

\section{Structure}

\begin{figure}
\includegraphics[width=8cm]{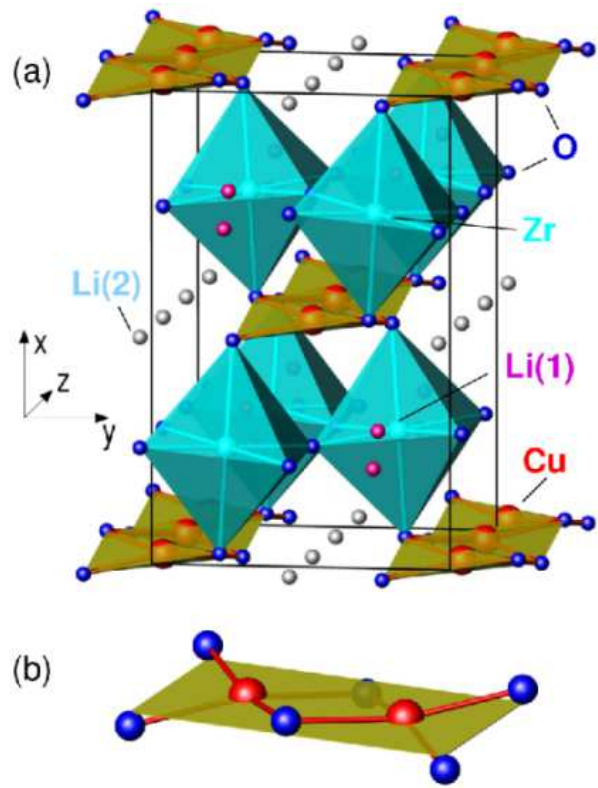}
\caption{\label{structure} (Color online) (a) Crystal structure of
  Li$_2$ZrCuO$_4$: Chains of edge-shared CuO$_4$ plaquettes (yellow)
  run along the $z$-direction and are connected by ZrO$_6$-octraheda
  (light blue). The fully occupied Li(2) site (gray) lies in the chain
  plane. The Li(1) split position (pink) alternates with the ZrO$_6$
  octahedra. (b) Section of the Cu-O chain: The O atoms of a
  plaquette are deflected from the mirror plane (shown in yellow).}
\end{figure}

The $\gamma$-phase of Li$_2$ZrCuO$_4$\cite{Duss02,footnote3} 
crystallizes in an orthorhombic space group and is shown in
Fig.~\ref{structure}a.  The compound contains distorted CuO$_4$
plaquettes which are arranged as edge-shared CuO$_2$ chains with an
Cu-O-Cu bond angle of 94.13$^\circ$. These edge shared CuO$_2$ chains
form layers together with the Li(2) atoms in the $yz$-plane
interconnected by ZrO$_6$ octrahedra.  The alternating arrangement of
the connecting ZrO$_6$ octrahedra lead to the deviation of the CuO$_2$
chains from the ideal planar geometry. In Fig.~\ref{structure}b
the deflection of the O atoms from the chain plane is illustrated in
detail.

As the CuO$_4$ plaquettes exhibit a localized, 
effective spin 1/2,
their linking and surrounding is of main importance for the formation
of the magnetic ground state.  A first structural analysis suggests
sizable ferromagnetic NN exchanges caused by the close to 90$^\circ$
Cu-O-Cu bond angle according to the Goodenough-Kanamori-Anderson 
rule, provided the direct Cu-Cu transfer integral can 
be ignored.\cite{GKA}
   
A further structural characteristic of the $\gamma$-phase is a split
position for Li(1) which is placed between the chain layers. The
distance of the Li(1) atoms from the high-symmetry
position\cite{footnote4} is 0.37\AA. While the influence of the split
position of Li(1) (corresponding to 50\% disorder within a classical
picture, ignoring possible tunneling processes of Li(1) between the
two sites of the split position) to the magnetic properties is
investigated carefully, we neglect the small disorder of 3-5 \% at the
Li(2) and Cu site in our theoretical calculations as a good
approximation.

It is well known that in the vicinity of a QCP small changes in any
parameter can be of crucial relevance for the ground state realized by
the system. Therefore, the above mentioned structural peculiarities of
Li$_2$ZrCuO$_6$ raise the following two questions especially related
to their influence on the electronic and magnetic properties.  (i)How
does the modeling of the Li(1) split position influences the
calculated ground state?  As there exists no standard procedure in
band structure codes to treat split positions 
 we suggest various
classical (i.e. within the adiabatic approximation) approaches. 
(ii) To which extend does the distortion of the
CuO$_4$ plaquettes influence the placement of the compound in the
$J_1-J_2$ phase diagram? Therefore a fictitious structure with ideal planar
chain geometry was also considered.\cite{footnote5}

\section{Results and Discussion}
\subsection{Modeling the Li(1) split position}

The vicinity of Li$_2$ZrCuO$_4$ to a QCP requires a careful check of
the structural description of the compound as a basis for our
theoretical calculation. Therefore the treatment of the Li(1) split
position might be crucial, although the different models may only
exhibit subtle differences. In the following we suggest different
models to handle the split position and discuss the results with
respect to the influence of these model assumptions on 
the relevant states and related dispersions.

\begin{figure}
\includegraphics[width=8cm]{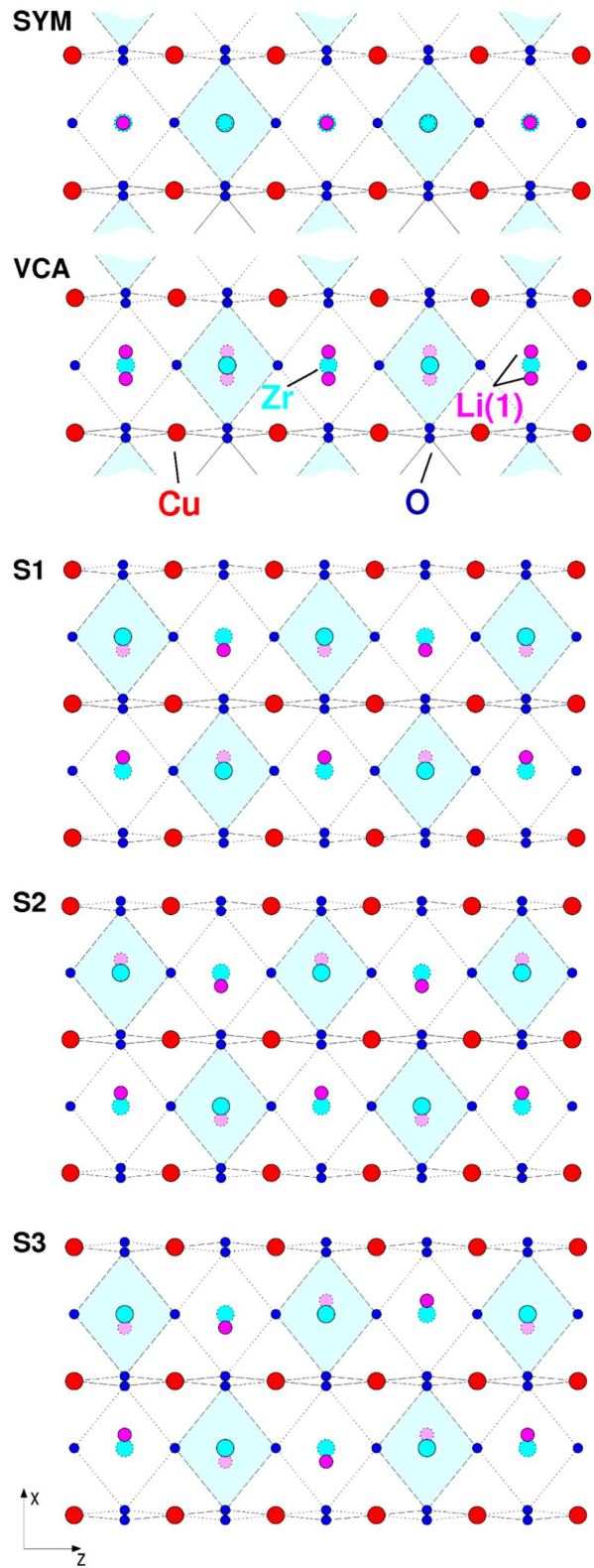}
\caption{\label{li-models} (Color online) Different structural models 
  to treat the Li(1) split position. From top to bottom:  SYM - high-symmetry 
position for Li(1), VCA - both sites occupied by partial Li(1) atoms, 
S1 to S3 - super-cells with different Li(1) pattern.
In S1 and S2 the Li(1) atoms along the chains are shifted parallel while 
in S3 the Li(1) atoms alter along the chain direction. S1 and S2 differ in the
 Li(1) inter-chain arrangement.  
The slight tilting of the ZrO$_6$-octrahedra
  (see Fig.~\ref{structure}) is not depicted for the sake of
  simplicity (full lines foreground, dashed lines background).}
\end{figure}

(i) To start from the most simplest approximation we performed
electronic structure calculation for Li(1) placed at the high-symmetry
position (see Fig.~\ref{li-models} (SYM)).

(ii) As an alternative structural model we carried out VCA
calculations (see Fig.~\ref{li-models} (VCA)). This approach occupies
both sites of the split position simultaneously by a half of a Li
atom. Each of this modified Li atoms is constructed by carrying the
half number of valence electrons and a modified core to conserve
charge.  For the VCA calculations the distance between the two sites
of the split position has to be enlarged to suppress the overlap of
the Li valence states and ensure the convergence of the
calculation.\cite{footnote6} Also asymmetric charge distributions between
the two sites of a split position (as e.g. 0.8 and 0.2 Li) lead to
nearly identical dispersion of the relevant states as long as the
total charge of one Li atom per split position is conserved. However,
deviations of the total charge of a split position cause sizable
changes in the electronic structure.

This is also the reason why the description of the Li(1) split
position by CPA fails, caused by the loss of charge balance at the
split position in the later approximation. In contrast to the VCA,
where the total charge at each split position sums up to one Li atom,
the random occupation of the Li(1) split sites in the CPA also
generates cases with empty and double occupied Li(1) sites with
dramatic influence on the band structure and the total energy.

(iii) As both latter approaches do not touch the local symmetry, within a
further model super-cells with different Li(1) patterns have been
calculated, breaking the local symmetry at the split position but
suffer from long range order (see Fig.~\ref{li-models} S1-S3).  In the
two upper panels the super-cells were constructed by a reduction of the space 
group
symmetry. The Li(1) atoms along the chains are all shifted in the same
direction while the Li(1) atoms of neighboring chains are parallel (S1)
or anti-parallel (S2) shifted.  The third super-cell
(Fig.~\ref{li-models} S3) shows an alternating displacement of the Li(1)
atoms along the chain obtained by a doubling of the cell in
$y$-direction.

\begin{figure}
\includegraphics[width=8.0cm, angle=0]{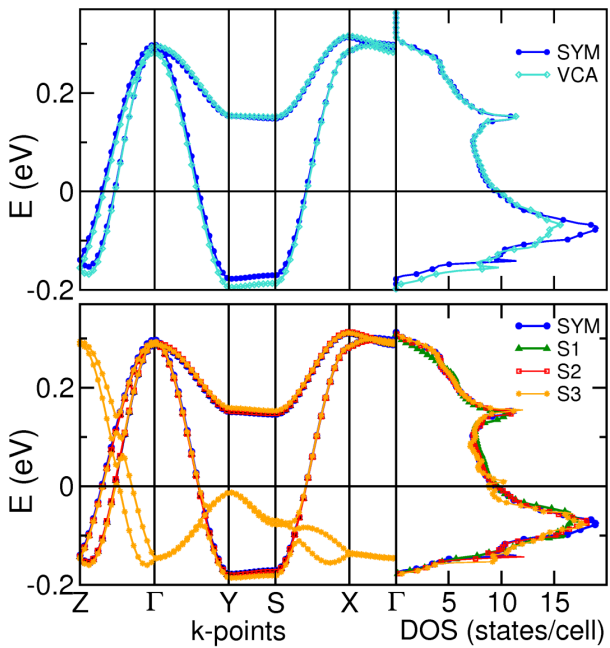}
\caption{\label{compared_bandsli} (Color online) Calculated LDA
  density of states and band structure for the physical relevant
  states using different structural models for the Li split position.
The comparison shows just minor differences.}
\end{figure}

The resulting bands and density of states for the different structural
models show a very similar total behavior (see
Fig.~\ref{compared_bandsli}).  Small deviations in VCA calculations can
be observed in the bonding region of the valence band around -5.5 and
-2.0 eV (not shown) in comparison to the high symmetry model. These
differences can be understood by the influence of the two modified Li
cores to the crystal potential and the corresponding shift of Li-O
states.
  
Comparing the density of states and LDA band structure for the 
anti-bonding $dp\sigma^*$ states, which are relevant for the magnetic
exchanges and therefore the magnetism of the system, we found only
tiny differences.  (i) The VCA model results in an slightly larger
bandwidth compared to the other models.  (ii) The high symmetry and
super-cell models nearly fit perfect. The additional bands of the S3
super-cell with an alternating Li pattern can be understood due to the
doubling of the primitive cell having its origin in the symmetry
breaking of the Li pattern.

\subsection{Electronic Structure and Magnetic Properties}\label{chap_elec}

As the different structural models for the Li(1) split position show
no significant difference in behavior for the relevant low energy states
(see Fig.~\ref{compared_bandsli}), the electronic structure and the
microscopic magnetic model are discussed for the SYM model as a good
representative for all different structural models. To estimate more
quantitatively the differences caused by the choice of the
structural model the results for the SYM and the VCA models are compared.

\begin{figure}[h]
\includegraphics[width=7.5cm,angle=0]{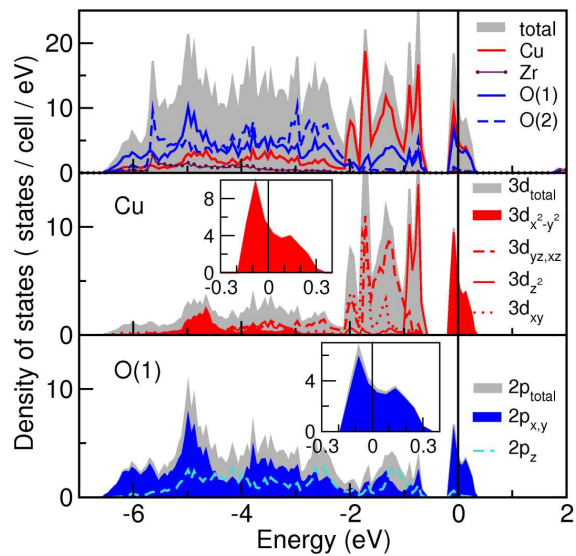}
\caption{\label{dos} 
(Color online) Calculated atom and orbital resolved density of states (LDA) for
Li$_2$ZrCuO$_4$ applying the SYM model for the Li(1) split position
(see text). The Li contribution to the valence band is negligible. Cu
and O dominate the anti-bonding $dp\sigma^*$ band.}
\end{figure}

In Fig.~\ref{dos} the atom and orbital resolved density of states for
Li$_2$ZrCuO$_4$ are shown.  The valence band has a width of about 7eV,
comparable with other chain cuprates and is dominated by Cu and O
states, especially at the lower edge of the valence band around -6 eV
(bonding states) and at the energy range close to zero
(anti-bonding states). States of the Zr-O octrahedra
appear in the middle part of the valence band. Their contribution to
the states close to the Fermi energy is negligible.
\begin{figure}
\includegraphics[width=5.0cm,angle=270]{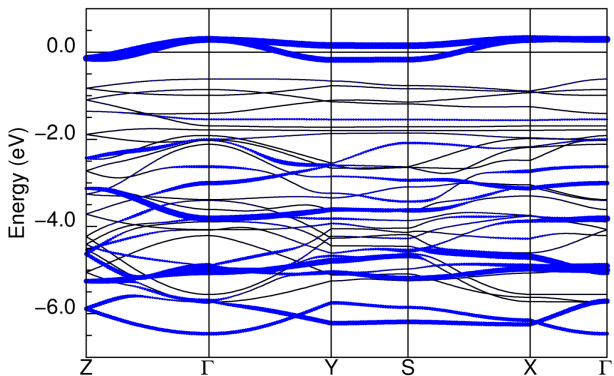}
\caption{\label{fatbands}(Color online) Valence band together with the band
  characters of the Cu $3d$ and O $2p$ anti-bonding states (blue) indicated 
by the line width. The band
  at Fermi energy is dominated by Cu and O anti-bonding $dp\sigma^*$
  states.}
\end{figure}

The metallic character of the LDA results is in contradiction to the
experimentally observed insulating ground state. This effect is a well
known shortcoming of the LDA as it underestimates the strong
correlations in the Cu $3d$ states.  The related strong Coulomb
repulsion on Cu sites can be considered explicitly by a model approach
(e.g. a single-band Hubbard model using tight-binding parameters
derived from the LDA) or by LDA+$U$ calculations.  Both methods were
applied in this study and lead to an insulating ground state in
agreement with the experimental data.  Nevertheless, the LDA
calculation delivers important insights into the electronic structure
and provides the relevant bands for the development of a
microscopically based picture for the magnetic ground state of
Li$_2$ZrCuO$_4$.

The well separated states around the Fermi energy are of main interest
as they determine the magnetism of the system.  These states are
dominated by the in-plane orbitals of the CuO$_4$ plaquettes
corresponding to the anti-bonding $dp\sigma^*$ states (see two lower
panels in Fig.~\ref{dos}).  The distortion of the plaquettes from an
ideal planar chain geometry does not affect significantly the band
characters of the anti-bonding $dp\sigma^*$ states. The hybridization
with out-of-plane states is very small and does not differ from
compounds with ideal planar CuO$_2$ chains.

\begin{figure}
\includegraphics[width=8.0cm, angle=0]{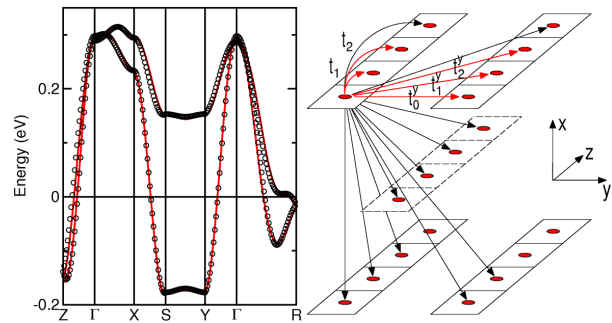}
\caption{\label{bandfit}(Color online) Left: Comparison of the calculated band
  structure (black dots) and the least-square fit (red line) of the TB
  model. Right: Considered hoppings for the TB model. The leading
  interactions are labeled and highlighted by red colors. }
\end{figure}

In Fig.~\ref{fatbands} the band structure of the valence band together
with the Cu 3$d_{x^2-y^2}$ and O 2$p_{x,y}$ characters are shown. The
anti-bonding $dp\sigma^*$ bands (see Fig.~\ref{bandfit} left) show
their main dispersion along the chain direction (Z-$\Gamma$) and in
the chain plane ($\Gamma$-Y, S-X) indicating the leading magnetic
exchanges in the layers. Perpendicular to these directions the
dispersion is small, suggesting small exchanges between the chain
layers.

To unravel the hierarchy of exchanges in the system as well as to
estimate a cut-off for negligible long range exchanges in further
super-cell calculations, an effective one-band model was fitted to the
anti-bonding $dp\sigma^*$ bands. Effective hoppings for Cu-Cu
distances up to 12\,\AA\ are considered (see Fig.~\ref{bandfit} right)
and calculated using a least-square fit procedure. The resulting TB
fit and the LDA bands are in perfect agreement (see Fig.~\ref{bandfit}
left).

\begin {table}[tbh]
\begin{ruledtabular}
\begin{tabular}{l l l l l l}
$t_i/meV$ &  $t_1$ & $t_2$ & $t^{y}_{0}$ & $t^{y}_{1}$ & $t^{y}_{2}$ \\ 
\hline
SYM & -42 & -56 & -26 & 14 & -25\\
VCA & -44 & -55 & -25 & 15 & -26\\
FIC & -86 & -92 & -49 & 24 & -21\\
\end{tabular}
\caption{\label{tab_ts} Resulting hopping integrals from TB fits for different
  structural models.}
\end{ruledtabular}
\end{table}

All obtained transfer integrals larger than 5\,meV are summarized in 
Table~\ref{tab_ts}. 
We found the main hoppings between NN and NNN along 
the chains and three considerable inter-chain hoppings in the chain
plane.  Comparing the SYM results with results from the VCA model, we
see that the differences in the hopping integrals are very small and in
the same order as the accuracy of the TB fit procedure. This
indicates that the chosen structural model for the Li(1) split position
is irrelevant for the calculated coupling constants of Li$_2$CuZrO$_4$.
Taking into account the strong Coulomb repulsion in a model approach
we construct from our TB model a single-band Hubbard model 
(with a typical value \cite{Parmigiani} $U_{eff}= 4.2$\,eV for edge 
shared CuO$_2$ chains) which can be mapped subsequently onto a
Heisenberg model. We find $J_1^{AFM}= 1.7 \pm 0.1$\,meV for the NN
exchange and $J_2^{AFM}= 3.0 \pm 0.1$\,meV for the NNN exchange.\cite{footnoteerror}

The exchange parameters resulted from the effective TB model
disregard the FM contributions to the exchanges which are, however,
expected between NN plaquettes in the compound due to the close to
90$^{\circ}$ bond angle Cu-O-Cu along the chains. Naturally, these FM
interactions are included in LDA+$U$ calculations. To evaluate $J_1$
and $J_2$ on a quantitative level, we performed a series of super-cell
calculations. The comparison of the total energies of these
super-cells with different spin arrangements\cite{footnote7} and a
subsequent mapping onto a Heisenberg model leads to a FM NN exchange
$J_1^{TOT}$= -11.2 meV and an AFM NNN exchange $J_2^{TOT}$= 2.2\,meV. The ground
states of the compound is determined by the ratio
$\alpha=-J_2^{TOT}/J_1^{TOT}=0.22\pm0.03$ that would result in a FM ordered ground
state in close vicinity of a QCP ($\alpha=0.25$). The variation of
$U_{3d}$ from 5.5\,eV to 8.0\,eV in the LDA+$U$ calculation does not
change this behavior qualitatively, although for the smaller $U$
values a slightly increased $\alpha$ is obtained.\cite{footnoteSLD}  
Besides a possible
small error in the calculated exchange parameters, the experimentally
observed spiral state from NMR data\cite{Tarui08} might result due to
the sizable inter chain couplings $J^y_0 = 4(t^y_0)^2/U_\mathrm{eff}
\sim$ 0.6\,meV and $J^y_2 \sim$ 0.6\,meV that both stabilize long
range order.

\begin {table}
\begin{ruledtabular}
\begin{tabular}{l l l l l l l}
$J_i/meV$&$J_1^{TOT}$&$J_2^{TOT}$&$J_1^{AFM}$&$J_1^{FM}$&$J_2^{AFM}$&$J_2^{FM}$\\ 
\hline
SYM & -11.2 & 2.2 & 1.7 & -12.9 & 3.0 & -0.8\\
VCA & -10.6 & 2.5 & 1.8 & -12.4 & 2.9 & -0.4\\
FIC & -10.2 & 5.9 & 7.0 & -17.2 & 8.0 & -1.1\\
\end{tabular}
\end{ruledtabular}
\caption{\label{tab2} Total and partial exchange integrals for
  different structural models from TB model and LDA+$U$ calculation ($U_{3d}=7.5$\,eV).}
\end{table}

\subsection{Influence of chain-buckling}

\begin{figure}
\includegraphics[width=8cm,angle=0]{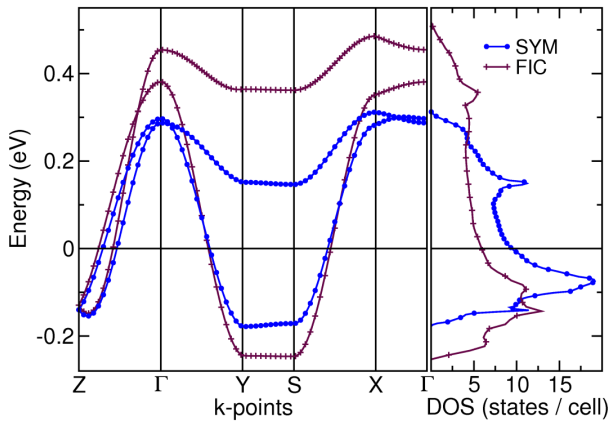}
\caption{\label{es_planar} (Color online)
Comparison of band structure and density of states at Fermi energy for both 
chain geometries. The small displacement of O atoms from the mirror plane leads to a 
decrease of the band dispersion.}
\end{figure}

Whereas the different treatment of the Li(1) split position does
 not influence the band structure and the corresponding
magnetic model essentially, the role of the distortion of the CuO$_4$ plaquettes
(see Fig.~\ref{structure}b) remains to be addressed: How much does the
non-planar geometry of the chains influence the balance of the NN,
NNN and inter-chain couplings and the resulting placement of the
compound in the $J_1$-$J_2$ phase diagram, especially its vicinity to
the QCP? For a fictitious planar chain,\cite{footnote5} changes are
expected since a changed Cu-O-Cu bond angle (changed by about
0.8$^\circ$) will influence J$_1^{TOT}$ and the different orientation of the
anti-bonding $pd\sigma^*$ ``molecular orbital'' will alter $J_2^{TOT}$. Since
these changes are hardly predictable quantitatively from structural
considerations only, we apply our mapping approach also to a
fictitious compound with ideal planar CuO$_2$ chains.

In Fig.~\ref{es_planar} the resulting anti-bonding $dp\sigma^*$ bands
and the related density of states for this fictitious structure are
shown.  For a planar chain geometry we obtain an about 50\% larger
bandwidth, but a very similar shape of the bands compared with the
real structure. On first glance, this suggests a scaling of the
leading hopping parameters, only. Our TB fit yields nearly a doubling
of the hopping integrals as listed in Tab.~\ref{tab_ts}, thus a
dramatic change of the corresponding exchange terms can be
expected. To take into account the large FM contributions to $J_1^{TOT}$, we
carried out LDA+$U$ calculations for the same super-cells used
before.\cite{footnote7}

Surprisingly, we find a nearly unchanged NN exchange J$_1^{TOT}$, but
only due to the compensation of the increased individual contributions
$J^{FM}_1$ and $J^{AFM}_1$ (see Tab.~\ref{tab2}). Since the FM
contribution $J^{FM}_2$ to $J^{TOT}_2$ remained small - as for the
real structure - it basically scales like $J^{AFM}_2 \sim
4t_2/U_{\mathrm{eff}}$. Thus, we obtain a large change in $J^{TOT}_2$
and consequently in the frustration ratio $\alpha \sim 0.6$ for the
planar chains.  In consequence, the ratio $\alpha =-J_2/J_1$ depends
strongly on structural details of the local Cu-O environment. This is
especially important regarding the vicinity of the system to the QCP.
On the other hand this sensitivity of $\alpha$ to the chain buckling
may provide the opportunity of manipulating the ground state of the
system selectively by substitution or external pressure.

Unfortunately, a planar chain arrangement leads also to a sizable
increase of the inter chain coupling. This can be understood as the
decrease of buckling increases the inter-chain overlap of the O
orbitals belonging to the $pd\sigma^*$ states. This way, a tendency
towards long range order is stabilized, although the quantitative
influence of inter-chain couplings to the ground state is sparsely
considered in the literature.

\subsection{Stability of the Li(1) split position}\label{Li_energychp}

\begin{figure}
\includegraphics[width=8cm]{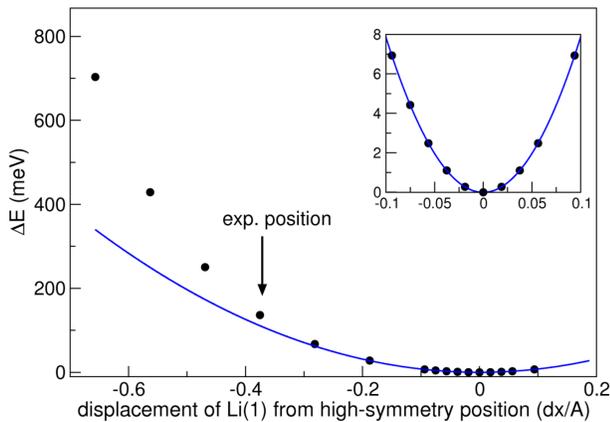}
\caption{\label{li_ee} (Color online) Energy difference depending on 
the displacement of the Li(1) atom from the high-symmetry
position. The experimental observed displacement of 0.4~\AA\ is
indicated by an arrow. The inset shows a harmonic (second order) fit
to the calculated energy differences for small displacements from the
equilibrium position ($\pm$0.1\AA). The same fit is shown in the main
panel, indicating deviations from the harmonic behavior at larger
Li(1) elongations.  }
\end{figure}

Although the different treatments of the Li(1) split position yield no
significant influence on the magnetic exchange parameters the split position 
may
influence other properties of the compound. Related to the split
position, the local symmetry or disorder could modulate the pitch
angle of the helical state or be crucial for the formation of a
possible multiferroic phase.  Furthermore, $^7$Li NMR and complex
dielectric measurements reveal a glass-like ordering of the Li(1) ions
below $T_g=100$\,K.\cite{Vavilova08}  Thus, using our super-cells
simulating different types of Li(1) order (see Fig.~\ref{li-models}),
we tackled the questions whether a tendency to a static Li(1) order
can be supported by calculations. 

For the purpose of computational feasibility, we restricted ourselves to
a fourfold cell (32 atoms).\cite{footnote8} The comparison of the
total energy for the different ordered Li(1) patterns (SYM, S1, S2,
S3) favors energetically the high-symmetry model (SYM) with a single
Li(1) position. The second lowest in energy is the structure with
alternating Li(1) displacement along the chain direction (S3).  For
this Li(1) arrangement, the dependence of the total energy on the
displacement is depicted in Fig. \ref{li_ee}.  The resulting curve
does not indicate a double well potential as expected for a Li(1)
split position, but a minimum around the high symmetry position and a
harmonic behavior (see inset Fig.~\ref{li_ee}) up to almost 0.3\,\AA\
Li(1) displacement. According to our calculations, for the
experimentally suggested Li(1) split position (marked by the arrow in
Fig.~\ref{li_ee}) only slight anharmonic effects could be expected.
We also investigated whether the strong Coulomb repulsion on the Cu
site could be the reason for the discrepancy between the single Li(1)
position suggested by the calculations and the experimental split
position using LDA+$U$. With U$_{3d}$ in the range from 5 - 7.5\,eV we
do not obtain any significant changes of the position of the minimum
and the slope of the curve, suggesting that the correlations at the Cu
site play only a minor role for a possible Li(1) order.

So far our calculations with a fixed lattice (apart from Li(1)) can
not support the split position, but favor the high-symmetry solution
(SYM). The discrepancy with the experiment that observed a glass-like
Li(1) ordering below 100\,K, may have the following origins: (i)
dynamic effects that violate the adiabatic approximation used in the
calculations or (ii) a correlated Li(1)-O(1) position without long
range order that would be difficult to detect in X-ray diffraction
experiments. An investigation of these effects is beyond the scope of
the present study but should be carried out in future since it might
be crucial for the final picture of the compound.

\subsection{Cluster Calculations}
In order to provide an independent microscopic interpretation of the
exchange integrals reported above, we performed exact diagonalization
studies for the commonly used five-band extended $pd$ Hubbard model for
finite open Cu$_n$O$_{2n+2}$ clusters. The most reliable way to choose
a reasonable parameter set by fitting various spectroscopic
experiments cannot be used here due to the lack of single crystals and
available experimental data. Since no parameter set for the five-band
$pd$ model for a buckled chain geometry is available, we considered
only the fictitious planar case (FIC) without any O-buckling. This
way, a direct comparison to the band structure derived results (FIC)
should be more reliable. Thus, we started from an available parameter
set known for the well-studied closely related sister compound
Li$_2$CuO$_2$ with planar CuO$_2$ chains.\cite{Malek08,Malek1} Then,
only two Hamiltonian parameters - the mean Cu-O on-site difference
$\Delta_{pd}=3.5$~eV (3.75~eV) and $K_{pd} = 54$~meV (81~meV) - have
to be slightly lowered to obtain consistency with the band structure
results for the exchange integrals $J_1$ and $J_2$ reported above. For comparison the Li$_2$CuO$_2$ parameters are given in
brackets\cite{remark}. The slightly smaller $\Delta_{pd}$ is
in line with the slight increase of the Cu-O distance by about 2\%
along the CuO$_2$ chains for Li$_2$ZrCuO$_4$ compared to
Li$_2$CuO$_2$.

With these parameters the title compound exhibits a charge gap of
about 2.35~eV as expected for an undoped cuprate. Therefore all
eigenstates below that energy are spin states which can be described
also reasonably well by a Heisenberg Hamiltonian. In case of a
Cu$_5$O$_{10}$ cluster we deal with 10 spin states. Mapping for
instance these 10 spin states onto a 1D $J_1$-$J_2$-$J_3$-Heisenberg
model,\cite{footnote9} we arrive then at $J_1=-10.6$~meV, $J_2= 5.9$~meV and
$J_3=-0.3$~meV close to the values (FIC) given in Tab.~II.  Notice that
our mapping on larger clusters with 5 Cu-sites is considered to
be more accurate than the determination of $J_1$ from a
Cu$_2$O$_4$-dimer and $J_2$ from a Cu$_2$O$_8$ pseudo-trimer, where
the central Cu has been removed.\cite{Mizuno99}

In the context of the lower value of $J_1$ compared with 
Ref.~\onlinecite{Drechs07} we would like to mention that from a microscopic
point of view $J_1$ depends strongly on the adopted value of the
direct ferromagnetic Cu-O exchange parameter $K_{pd}$. Within
perturbation theory it occurs already in the second order whereas the
antiferromagnetic superexchange due to deviations from the 90$^\circ$
and/or due to different crystal fields seen by the O orbitals along
and perpendicular to the chain direction as well as the ferromagnetic
Hund's-rule coupling and the O-sites occur in the fourth order, 
only.\cite{Eskes93} Analytically and numerically one obtains a linear
dependence. Unfortunately, at present there is no consensus about a
reliable value of $K_{pd}$ based on microscopic
considerations. Therefore it is frequently treated as fitting
parameter to reproduce empirical or otherwise determined values of
$J_1$ just as done here.  For other CuO$_2$ edge-shared chain cuprates
$K_{pd}$-values in between 50~meV and 110~meV have been employed by
various authors.\cite{Mizuno99,Braden96,Drechs07,Malek08,Malek1} In
our previous work on Li$_2$ZrCuO$_4$
\cite{Drechs07} 70 meV has been adopted, thus with the given above 54~meV
we are near to the lower bound of so far used $K_{pd}$-values.  The
former value was based on 1D-fits of the magnetic susceptibility
$\chi(T)$ and of the magnetic specific heat $c_v(T)\approx c_p(T)$,
which might be significantly affected by the ignored inter-chain
interaction. Due to the vicinity to the critical point the calculated
1D saturation field at $T=0$\,K is very small of the order of 4 to
5~Tesla. Hence, the experimentally measured saturation field of about
30~Tesla (at low temperature) is clearly dominated by the inter-chain
interaction in accord with our calculated sizable inter-chain couplings
(see Sec.~\ref{chap_elec}).

\begin{figure}
\includegraphics[width=7cm,angle=-90]{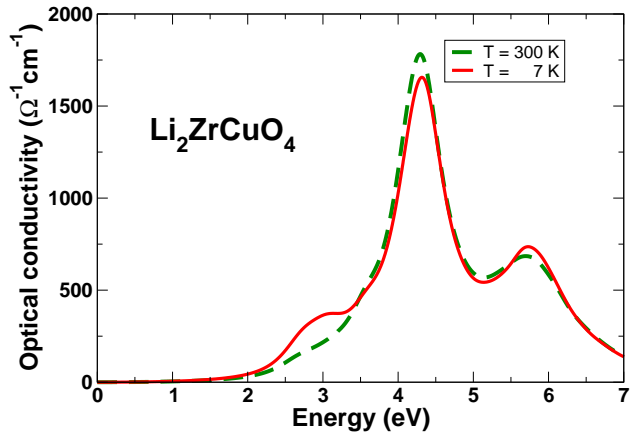}
\caption{\label{opt} (Color online) Optical conductivity of Li$_2$ZrCuO$_4$ 
within 
the five-band extended Hubbard model for a Cu$_5$O$_{12}$ cluster.}
\end{figure}

In order to stimulate optical measurements and for a further
experimental check of the applied Hamiltonian parameters we calculated
(see Fig.~\ref{opt}) the in-chain optical conductivity for 300~K and
7~K using the technique presented in Ref.~\onlinecite{Malek08}. We
draw attention to the strong temperature effect near 3\,eV between low
(7\,K) and room temperature (300\,K). This phenomenon is related to
the different optical response in different initial spin states. For
relatively small exchange integrals $J_1$ and $J_2$ the thermal
population of these spin states changes markedly in the mentioned
temperature range between zero and room temperature. Since the NN
exchange $J_1$ (FIC) is mostly unchanged compared to the buckled
structure (SYM, VCA), a strong temperature dependence of the optical
conductivity should hold for the real compound.

Further investigations to settle the value of $J_1$ are highly
desirable not only to get access to the Hamiltonian
parameters,\cite{footnote10} but also to check for renormalization
and creation of longer-ranged couplings due to strong enough
spin-phonon interaction.\cite{Weisse99} The strong coupling of the
exchange parameter $J_2$ to the O buckling, indicated by our band
structure calculations (see Tab.~\ref{tab2}), opens room for such
a scenario.

\section{Conclusions and Summary}

To summarize, applying LDA and LDA+$U$ band structure calculations as
well as exact diagonalization studies of an extended $pd$ Hubbard
model for large Cu$_n$O$_{2n+2}$ clusters (n=5), that were finally all
mapped onto a Heisenberg model, we obtain a consistent,
microscopically based picture: (i) Li$_2$ZrCuO$_4$ can be understood
as a quasi-1D chain compound with FM NN $J_1$ and AFM NNN $J_2$
exchange in close vicinity to a QCP.
(ii) We find sizable inter-chain couplings that
should be relevant for the magnetic ground state of the system,
especially due to the vicinity of the system to a critical point where
the impact of the leading exchange parameters on the ground state gets
small. (iii) Calculations for a fictitious structure with planar
chains indicate that the balance between $J_1$ and $J_2$ depends
heavily on structural details, especially on the magnitude of the
CuO$_2$ chain buckling. (iv) This also applies for the inter-chain
coupling which is strongly increased for the fictitious planar
geometry.

Our findings suggest that the buckling of the CuO$_2$ chain is crucial
for the vicinity to the QCP, therefore it should be possible to tune
the system towards that point by chemical substitution at the Zr site
or by external pressure. On the other hand, it would be desirable to
decrease the inter-chain coupling for which the orientation of the CuO
$pd\sigma$ orbitals is more relevant than inter-atomic distances
according to our results. Thus, a search for good quasi 1D systems
should turn towards crystal structures where the chains are arranged
in a strongly non-planar pattern.

Our simulations indicate that the Li(1) split position has only a
negligible direct influence on the spin system, assuming that all
other atoms of the structure are fixed to their experimental position.
In particular, our calculations do not support the experimental
observation of a glassy-like order for Li(1) on this split
position\cite{Vavilova08} since it yields a single Li(1) position as
most energetically favorable. However, an indirect influence of the
Li(1) position on the spin system is possible: Taking into account a
possible relaxation of the neighboring oxygen atom in the CuO$_2$
chain could result in a sizable modification of the exchange
parameters as shown by the strong influence of this O position on the
leading exchange terms. Another possible origin of the discrepancy
concerning the Li(1) position could arise from dynamic and
non-adiabatic effects that are beyond the scope of the present study.

\section{Acknowledgment}
We acknowledge the DFG (Emmy-Noether program, H.R.) and the Grant
DR269/3-1 (S.-L.D., J.M.), the German-Israel foundation (H.R.), and
the ASCR project AVOZ10100520 (J.M.) for financial support. The ZIH and
the IFW Dresden are acknowledged for the use of their computer
facilities. We thank O.~Janson, D.~Kasinathan, H.~Eschrig,
N.M.~Plakida, A.S.~Moskvin, and R.~Kuzian for fruitful discussions.


\begin{thebibliography}{31}
\expandafter\ifx\csname natexlab\endcsname\relax\def\natexlab#1{#1}\fi
\expandafter\ifx\csname bibnamefont\endcsname\relax
  \def\bibnamefont#1{#1}\fi
\expandafter\ifx\csname bibfnamefont\endcsname\relax
  \def\bibfnamefont#1{#1}\fi
\expandafter\ifx\csname citenamefont\endcsname\relax
  \def\citenamefont#1{#1}\fi
\expandafter\ifx\csname url\endcsname\relax
  \def\url#1{\texttt{#1}}\fi
\expandafter\ifx\csname urlprefix\endcsname\relax\def\urlprefix{URL }\fi
\providecommand{\bibinfo}[2]{#2}
\providecommand{\eprint}[2][]{\url{#2}}

\bibitem[{\citenamefont{McKenzie}(1997)}]{McKenzie}
\bibinfo{author}{\bibfnamefont{R.}~\bibnamefont{McKenzie}},
  \bibinfo{journal}{Sience} \textbf{\bibinfo{volume}{278}},
  \bibinfo{pages}{820} (\bibinfo{year}{1997}).

\bibitem[{\citenamefont{Misguich and Sindzingre}(2007)}]{Misguich}
\bibinfo{author}{\bibfnamefont{G.}~\bibnamefont{Misguich}} \bibnamefont{and}
  \bibinfo{author}{\bibfnamefont{P.}~\bibnamefont{Sindzingre}},
  \bibinfo{journal}{European Physical journal B} \textbf{\bibinfo{volume}{59}},
  \bibinfo{pages}{305} (\bibinfo{year}{2007}).

\bibitem[{\citenamefont{Johnston}(1997)}]{Johnston}
\bibinfo{author}{\bibfnamefont{D.}~\bibnamefont{Johnston}},
  \emph{\bibinfo{title}{Normal-state magnetic properties of single-layer
  cuprate high-temperature superconductors and related materials}},
  vol.~\bibinfo{volume}{10} (\bibinfo{publisher}{North-Holland},
  \bibinfo{year}{1997}).

\bibitem[{\citenamefont{Melzi et~al.}(2000)\citenamefont{Melzi, Carretta,
  Lascialfari, Mambrini, Troyer, Millet, and Mila}}]{Melzi00}
\bibinfo{author}{\bibfnamefont{R.}~\bibnamefont{Melzi}},
  \bibinfo{author}{\bibfnamefont{P.}~\bibnamefont{Carretta}},
  \bibinfo{author}{\bibfnamefont{A.}~\bibnamefont{Lascialfari}},
  \bibinfo{author}{\bibfnamefont{M.}~\bibnamefont{Mambrini}},
  \bibinfo{author}{\bibfnamefont{M.}~\bibnamefont{Troyer}},
  \bibinfo{author}{\bibfnamefont{P.}~\bibnamefont{Millet}}, \bibnamefont{and}
  \bibinfo{author}{\bibfnamefont{F.}~\bibnamefont{Mila}},
  \bibinfo{journal}{Phys. Rev. Lett.} \textbf{\bibinfo{volume}{85}},
  \bibinfo{pages}{1318} (\bibinfo{year}{2000}).

\bibitem[{\citenamefont{Melzi et~al.}(2001)\citenamefont{Melzi, Aldrovandi,
  Tedoldi, Carretta, Millet, and Mila}}]{Melzi01}
\bibinfo{author}{\bibfnamefont{R.}~\bibnamefont{Melzi}},
  \bibinfo{author}{\bibfnamefont{S.}~\bibnamefont{Aldrovandi}},
  \bibinfo{author}{\bibfnamefont{F.}~\bibnamefont{Tedoldi}},
  \bibinfo{author}{\bibfnamefont{P.}~\bibnamefont{Carretta}},
  \bibinfo{author}{\bibfnamefont{P.}~\bibnamefont{Millet}}, \bibnamefont{and}
  \bibinfo{author}{\bibfnamefont{F.}~\bibnamefont{Mila}},
  \bibinfo{journal}{Phys. Rev. B} \textbf{\bibinfo{volume}{64}},
  \bibinfo{pages}{024409} (\bibinfo{year}{2001}).

\bibitem[{\citenamefont{Rosner et~al.}(2002)\citenamefont{Rosner, Singh, Zheng,
  Oitmaa, Drechsler, and Pickett}}]{Rosner02}
\bibinfo{author}{\bibfnamefont{H.}~\bibnamefont{Rosner}},
  \bibinfo{author}{\bibfnamefont{R.}~\bibnamefont{Singh}},
  \bibinfo{author}{\bibfnamefont{W.}~\bibnamefont{Zheng}},
  \bibinfo{author}{\bibfnamefont{J.}~\bibnamefont{Oitmaa}},
  \bibinfo{author}{\bibfnamefont{S.-L.} \bibnamefont{Drechsler}},
  \bibnamefont{and} \bibinfo{author}{\bibfnamefont{W.}~\bibnamefont{Pickett}},
  \bibinfo{journal}{Phys. Rev Lett.} \textbf{\bibinfo{volume}{88}},
  \bibinfo{pages}{186405} (\bibinfo{year}{2002}).

\bibitem[{\citenamefont{Rosner et~al.}(2003)\citenamefont{Rosner, Singh, Zheng,
  Oitmaa, and Pickett}}]{Rosner_prb03}
\bibinfo{author}{\bibfnamefont{H.}~\bibnamefont{Rosner}},
  \bibinfo{author}{\bibfnamefont{R.}~\bibnamefont{Singh}},
  \bibinfo{author}{\bibfnamefont{W.}~\bibnamefont{Zheng}},
  \bibinfo{author}{\bibfnamefont{J.}~\bibnamefont{Oitmaa}}, \bibnamefont{and}
  \bibinfo{author}{\bibfnamefont{W.}~\bibnamefont{Pickett}},
  \bibinfo{journal}{Phys. Rev. B} \textbf{\bibinfo{volume}{67}},
  \bibinfo{pages}{014416} (\bibinfo{year}{2003}).

\bibitem[{\citenamefont{Bombardi et~al.}(2004)\citenamefont{Bombardi,
  Rodriguez-Carvajal, Matteo, de~Bergevin, Paolasini, Carretta, Millet, and
  Caciuffo}}]{Bombardi}
\bibinfo{author}{\bibfnamefont{A.}~\bibnamefont{Bombardi}},
  \bibinfo{author}{\bibfnamefont{J.}~\bibnamefont{Rodriguez-Carvajal}},
  \bibinfo{author}{\bibfnamefont{S.}~\bibnamefont{Matteo}},
  \bibinfo{author}{\bibfnamefont{F.}~\bibnamefont{de~Bergevin}},
  \bibinfo{author}{\bibfnamefont{L.}~\bibnamefont{Paolasini}},
  \bibinfo{author}{\bibfnamefont{P.}~\bibnamefont{Carretta}},
  \bibinfo{author}{\bibfnamefont{P.}~\bibnamefont{Millet}}, \bibnamefont{and}
  \bibinfo{author}{\bibfnamefont{R.}~\bibnamefont{Caciuffo}},
  \bibinfo{journal}{Phys. Rev. Lett.} \textbf{\bibinfo{volume}{93}},
  \bibinfo{pages}{027202} (\bibinfo{year}{2004}).

\bibitem[{\citenamefont{Shpanchenko et~al.}(2004)\citenamefont{Shpanchenko,
  Chernaya, Tsirlin, Chizhov, Sklovsky, Antipov, Khlybov, Pomjakushin,
  Balagurov, Medvedeva et~al.}}]{Shpan04}
\bibinfo{author}{\bibfnamefont{R.}~\bibnamefont{Shpanchenko}},
  \bibinfo{author}{\bibfnamefont{V.}~\bibnamefont{Chernaya}},
  \bibinfo{author}{\bibfnamefont{A.~A.} \bibnamefont{Tsirlin}},
  \bibinfo{author}{\bibfnamefont{P.}~\bibnamefont{Chizhov}},
  \bibinfo{author}{\bibfnamefont{D.}~\bibnamefont{Sklovsky}},
  \bibinfo{author}{\bibfnamefont{E.}~\bibnamefont{Antipov}},
  \bibinfo{author}{\bibfnamefont{E.}~\bibnamefont{Khlybov}},
  \bibinfo{author}{\bibfnamefont{V.}~\bibnamefont{Pomjakushin}},
  \bibinfo{author}{\bibfnamefont{A.}~\bibnamefont{Balagurov}},
  \bibinfo{author}{\bibfnamefont{J.}~\bibnamefont{Medvedeva}},
  \bibnamefont{et~al.}, \bibinfo{journal}{Chem. Mater.}
  \textbf{\bibinfo{volume}{16}}, \bibinfo{pages}{3267} (\bibinfo{year}{2004}).


\bibitem[{\citenamefont{Oka et~al.}(2008)\citenamefont{K. Oka, I. Yamada, M. Azuma, S. Takeshita, K. H. Satoh, A. Koda, R. Kadono, M. Takano and Y. Shimakawa}}]{Oka08}
\bibinfo{author}{\bibfnamefont{K.} \bibnamefont{Oka}},
  \bibinfo{author}{\bibfnamefont{I.}~\bibnamefont{Yamada}},
  \bibinfo{author}{\bibfnamefont{M.}~\bibnamefont{Azuma}},
\bibinfo{author}{\bibfnamefont{S.}~\bibnamefont{Takeshita}},
\bibinfo{author}{\bibfnamefont{K. H.}~\bibnamefont{Satoh}},
\bibinfo{author}{\bibfnamefont{A.}~\bibnamefont{Koda}},
\bibinfo{author}{\bibfnamefont{R.}~\bibnamefont{Kadono}},
\bibinfo{author}{\bibfnamefont{M.}~\bibnamefont{Takano}},
    \bibnamefont{and} \bibinfo{author}{\bibfnamefont{Y.}~\bibnamefont{Shimakawa}},
  \bibinfo{journal}{Inorg. Chem.} \textbf{\bibinfo{volume}{47}},
  \bibinfo{pages}{7355} (\bibinfo{year}{2008}).



\bibitem{Nitzsche} U.~Nitzsche, S.-L.~Drechsler and H.~Rosner, to be published.

\bibitem[{rem()}]{footnote1}
\bibinfo{note}{
Recent inelastic neutron scattering data by W. Lorenz {\it et al.} 
(privat communication) yield  $\alpha = 0.31$ for Li$_2$CuO$_2$ and
suggest that the ferromagnetic in-chain ordering in this compound
is stabilized by the antiferromagnetic inter-chain interactions.
}

\bibitem[{\citenamefont{Gippius et~al.}(2004)\citenamefont{Gippius, Morozova,
  Moskvin, Zalessky, Bush, Baenitz, Rosner, and Drechsler}}]{Gippi04}
\bibinfo{author}{\bibfnamefont{A.~A.} \bibnamefont{Gippius}},
  \bibinfo{author}{\bibfnamefont{E.~N.} \bibnamefont{Morozova}},
  \bibinfo{author}{\bibfnamefont{A.~S.} \bibnamefont{Moskvin}},
  \bibinfo{author}{\bibfnamefont{A.~V.} \bibnamefont{Zalessky}},
  \bibinfo{author}{\bibfnamefont{A.~A.} \bibnamefont{Bush}},
  \bibinfo{author}{\bibfnamefont{M.}~\bibnamefont{Baenitz}},
  \bibinfo{author}{\bibfnamefont{H.}~\bibnamefont{Rosner}}, \bibnamefont{and}
  \bibinfo{author}{\bibfnamefont{S.-L.} \bibnamefont{Drechsler}},
  \bibinfo{journal}{Phys. Rev. B} \textbf{\bibinfo{volume}{70}},
  \bibinfo{pages}{020406} (\bibinfo{year}{2004}).

\bibitem[{\citenamefont{Drechsler et~al.}(2005)\citenamefont{Drechsler,
  M\'{a}lek, Richter, Moskvin, Gippius, and Rosner}}]{Drechs05}
\bibinfo{author}{\bibfnamefont{S.-L.} \bibnamefont{Drechsler}},
  \bibinfo{author}{\bibfnamefont{J.}~\bibnamefont{M\'{a}lek}},
  \bibinfo{author}{\bibfnamefont{J.}~\bibnamefont{Richter}},
  \bibinfo{author}{\bibfnamefont{A.~S.} \bibnamefont{Moskvin}},
  \bibinfo{author}{\bibfnamefont{A.~A.} \bibnamefont{Gippius}},
  \bibnamefont{and} \bibinfo{author}{\bibfnamefont{H.}~\bibnamefont{Rosner}},
  \bibinfo{journal}{Phys. Rev. Lett.} \textbf{\bibinfo{volume}{94}},
  \bibinfo{pages}{039705} (\bibinfo{year}{2005}).

\bibitem[{\citenamefont{Enderle et~al.}(2005)\citenamefont{Enderle, Mukherjee,
  Fak, Kremer, Broto, Rosner, Drechsler, Richter, Malek, Prokofiev
  et~al.}}]{Enderle}
\bibinfo{author}{\bibfnamefont{M.}~\bibnamefont{Enderle}},
  \bibinfo{author}{\bibfnamefont{C.}~\bibnamefont{Mukherjee}},
  \bibinfo{author}{\bibfnamefont{B.}~\bibnamefont{Fak}},
  \bibinfo{author}{\bibfnamefont{R.}~\bibnamefont{Kremer}},
  \bibinfo{author}{\bibfnamefont{J.}~\bibnamefont{Broto}},
  \bibinfo{author}{\bibfnamefont{H.}~\bibnamefont{Rosner}},
  \bibinfo{author}{\bibfnamefont{S.-L.} \bibnamefont{Drechsler}},
  \bibinfo{author}{\bibfnamefont{J.}~\bibnamefont{Richter}},
  \bibinfo{author}{\bibfnamefont{J.}~\bibnamefont{Malek}},
  \bibinfo{author}{\bibfnamefont{A.}~\bibnamefont{Prokofiev}},
  \bibnamefont{et~al.}, \bibinfo{journal}{Europhys. Lett.}
  \textbf{\bibinfo{volume}{70}}, \bibinfo{pages}{237} (\bibinfo{year}{2005}).

\bibitem[{\citenamefont{Masuda et~al.}(2004)\citenamefont{Masuda, Zheludev,
  Bush, Markina, and Vasiliev}}]{Masuda}
\bibinfo{author}{\bibfnamefont{T.}~\bibnamefont{Masuda}},
  \bibinfo{author}{\bibfnamefont{A.}~\bibnamefont{Zheludev}},
  \bibinfo{author}{\bibfnamefont{A.}~\bibnamefont{Bush}},
  \bibinfo{author}{\bibfnamefont{M.}~\bibnamefont{Markina}}, \bibnamefont{and}
  \bibinfo{author}{\bibfnamefont{A.}~\bibnamefont{Vasiliev}},
  \bibinfo{journal}{Phys. Rev. Lett.} \textbf{\bibinfo{volume}{92}},
  \bibinfo{pages}{77201} (\bibinfo{year}{2004}).

\bibitem[{\citenamefont{Capogna et~al.}(2005)\citenamefont{Capogna, Mayr,
  Horsch, Raichle, Kremer, Sonin, and Keimer}}]{Capogna}
\bibinfo{author}{\bibfnamefont{L.}~\bibnamefont{Capogna}},
  \bibinfo{author}{\bibfnamefont{M.}~\bibnamefont{Mayr}},
  \bibinfo{author}{\bibfnamefont{P.}~\bibnamefont{Horsch}},
  \bibinfo{author}{\bibfnamefont{M.}~\bibnamefont{Raichle}},
  \bibinfo{author}{\bibfnamefont{R.}~\bibnamefont{Kremer}},
  \bibinfo{author}{\bibfnamefont{M.}~\bibnamefont{Sonin}}, \bibnamefont{and}
  \bibinfo{author}{\bibfnamefont{B.}~\bibnamefont{Keimer}},
  \bibinfo{journal}{Phys. Rev. B} \textbf{\bibinfo{volume}{71}},
  \bibinfo{pages}{140402} (\bibinfo{year}{2005}).

\bibitem[{\citenamefont{Drechsler et~al.}(2006)\citenamefont{Drechsler,
  Richter, Gippius, Vasiliev, Bush, Moskvin, Prots, Schnelle, and
  Rosner}}]{Drechs06}
\bibinfo{author}{\bibfnamefont{S.-L.} \bibnamefont{Drechsler}},
  \bibinfo{author}{\bibfnamefont{J.}~\bibnamefont{Richter}},
  \bibinfo{author}{\bibfnamefont{A.}~\bibnamefont{Gippius}},
  \bibinfo{author}{\bibfnamefont{A.}~\bibnamefont{Vasiliev}},
  \bibinfo{author}{\bibfnamefont{A.}~\bibnamefont{Bush}},
  \bibinfo{author}{\bibfnamefont{A.}~\bibnamefont{Moskvin}},
  \bibinfo{author}{\bibfnamefont{Y.}~\bibnamefont{Prots}},
  \bibinfo{author}{\bibfnamefont{W.}~\bibnamefont{Schnelle}}, \bibnamefont{and}
  \bibinfo{author}{\bibfnamefont{H.}~\bibnamefont{Rosner}},
  \bibinfo{journal}{Europhys. Lett.} \textbf{\bibinfo{volume}{73}},
  \bibinfo{pages}{83} (\bibinfo{year}{2006}).

\bibitem[{\citenamefont{Hase et~al.}(1993)\citenamefont{Hase, I.Terasaki, and
  Uchinokura}}]{Hase}
\bibinfo{author}{\bibfnamefont{M.}~\bibnamefont{Hase}},
  \bibinfo{author}{\bibnamefont{I.Terasaki}}, \bibnamefont{and}
  \bibinfo{author}{\bibfnamefont{K.}~\bibnamefont{Uchinokura}},
  \bibinfo{journal}{Phys. Rev. Lett.} \textbf{\bibinfo{volume}{70}},
  \bibinfo{pages}{3651} (\bibinfo{year}{1993}).

\bibitem[{\citenamefont{Drechsler et~al.}(2007)\citenamefont{Drechsler,
  Volkova, Vasiliev, Tristan, Richter, Schmitt, Rosner, M\'{a}lek, Klinger,
  Zvyagin et~al.}}]{Drechs07}
\bibinfo{author}{\bibfnamefont{S.-L.} \bibnamefont{Drechsler}},
  \bibinfo{author}{\bibfnamefont{O.}~\bibnamefont{Volkova}},
  \bibinfo{author}{\bibfnamefont{A.~N.} \bibnamefont{Vasiliev}},
  \bibinfo{author}{\bibfnamefont{N.}~\bibnamefont{Tristan}},
  \bibinfo{author}{\bibfnamefont{J.}~\bibnamefont{Richter}},
  \bibinfo{author}{\bibfnamefont{M.}~\bibnamefont{Schmitt}},
  \bibinfo{author}{\bibfnamefont{R.}~\bibnamefont{Rosner}},
  \bibinfo{author}{\bibfnamefont{J.}~\bibnamefont{M\'{a}lek}},
  \bibinfo{author}{\bibfnamefont{R.}~\bibnamefont{Klinger}},
  \bibinfo{author}{\bibfnamefont{A.~A.} \bibnamefont{Zvyagin}},
  \bibnamefont{et~al.}, \bibinfo{journal}{Phys. Rev. Lett.}
  \textbf{\bibinfo{volume}{98}}, \bibinfo{pages}{077202}
  (\bibinfo{year}{2007}).

\bibitem[{\citenamefont{Dussarrat et~al.}(2002)\citenamefont{Dussarrat, Mather,
  Caignaert, Domeng\`{e}s, Fletcher, and West}}]{Duss02}
\bibinfo{author}{\bibfnamefont{C.}~\bibnamefont{Dussarrat}},
  \bibinfo{author}{\bibfnamefont{G.~C.} \bibnamefont{Mather}},
  \bibinfo{author}{\bibfnamefont{V.}~\bibnamefont{Caignaert}},
  \bibinfo{author}{\bibfnamefont{B.}~\bibnamefont{Domeng\`{e}s}},
  \bibinfo{author}{\bibfnamefont{J.~G.} \bibnamefont{Fletcher}},
  \bibnamefont{and} \bibinfo{author}{\bibfnamefont{A.~R.} \bibnamefont{West}},
  \bibinfo{journal}{Journal of Solid State Chemistry}
  \textbf{\bibinfo{volume}{166}}, \bibinfo{pages}{311} (\bibinfo{year}{2002}).

\bibitem{footnote2} The total energy calculations
for the Li(1) position (Section~\ref{Li_energychp}) have been carried
out with both versions to ensure the independence from the basis set.

\bibitem[{\citenamefont{Koepernik and Eschrig}(1999)}]{Koep1}
\bibinfo{author}{\bibfnamefont{K.}~\bibnamefont{Koepernik}} \bibnamefont{and}
  \bibinfo{author}{\bibfnamefont{H.}~\bibnamefont{Eschrig}},
  \bibinfo{journal}{Phys. Rev. B} \textbf{\bibinfo{volume}{59}},
  \bibinfo{pages}{1743} (\bibinfo{year}{1999}).

\bibitem[{\citenamefont{Perdew and Wang}(1992)}]{PerdW}
\bibinfo{author}{\bibfnamefont{J.}~\bibnamefont{Perdew}} \bibnamefont{and}
  \bibinfo{author}{\bibfnamefont{Y.}~\bibnamefont{Wang}},
  \bibinfo{journal}{Phys. Rev. B} \textbf{\bibinfo{volume}{45}},
  \bibinfo{pages}{13244} (\bibinfo{year}{1992}).

\bibitem[{\citenamefont{Perdew, Burke and Ernzerhof}(1996)}]{Burke96}
\bibinfo{author}{\bibfnamefont{J.}~\bibnamefont{Perdew}},
 \bibinfo{author}{\bibfnamefont{K.}~\bibnamefont{Burke}},
\bibnamefont{and}
  \bibinfo{author}{\bibfnamefont{M.}~\bibnamefont{Ernzerhof}},
  \bibinfo{journal}{Phys. Rev. Lett.} \textbf{\bibinfo{volume}{77}},
  \bibinfo{pages}{3865} (\bibinfo{year}{1996}).


\bibitem[{\citenamefont{Eschrig et~al.}(2003)\citenamefont{Eschrig, Koepernik,
  and Chaplygin}}]{Esch1}
\bibinfo{author}{\bibfnamefont{H.}~\bibnamefont{Eschrig}},
  \bibinfo{author}{\bibfnamefont{K.}~\bibnamefont{Koepernik}},
  \bibnamefont{and}
  \bibinfo{author}{\bibfnamefont{I.}~\bibnamefont{Chaplygin}},
  \bibinfo{journal}{J. Solid States Chemistry} \textbf{\bibinfo{volume}{176}},
  \bibinfo{pages}{482} (\bibinfo{year}{2003}).

\bibitem {footnote3}Since in Ref.~\onlinecite{Duss02} and in the ICSD(59618) the lattice
  parameters and internal coordinates are inconsistent or erroneous,
  we explicitly provide the parameters used in the calculations: Space
  group Cccm (66), $a=9.385$\,\AA, $b=5.895$\,\AA, $c=5.863$\,\AA , Cu(0.0, 0.0,
  0.5), Li(1)(0.21, 0.75, 0.0), Li(2)(0.0, 0.5, 0.25), Zr(0.25, 0.25,
  0.0), O(1)(-0.0246, 0.228, 0.5), O(2)(0.2662, 0.0, 0.25).

\bibitem[{\citenamefont{Kanamori}(1957)}]{GKA}
\bibinfo{author}{\bibfnamefont{J.}~\bibnamefont{Kanamori}},
  \bibinfo{journal}{Theor. Phys. (Kyoto)} \textbf{\bibinfo{volume}{17}},
  \bibinfo{pages}{177} (\bibinfo{year}{1957}).

\bibitem{footnote4}For the high-symmetry position (SYM) Li(1) is placed at
  (0.25, 0.75, 0.0).

\bibitem{footnote5}O(1) was placed at (0.0, 0.228, 0.5). This structural
  variation decreases the Cu-O bond length by 0.01\AA$\,$ and increases the
  Cu-O-Cu bond angle along the edge shared chain by 0.8$^\circ$.

\bibitem{footnote6}The experimentally determined distance between the two
  possible sites of a split position is 0.757\,\AA. For the VCA
  calculation the distance has been enlarged to 1.314\,\AA.

\bibitem[{\citenamefont{Parmigiani et~al.}(1996)\citenamefont{Parmigiani, 
Sangaletti, Goldoni, del Pennino, Kim, Shen, Revcolevschi and Dhalenne}}]{Parmigiani}
\bibinfo{author}{\bibfnamefont{F.}~\bibnamefont{Parmigiani}},
  \bibinfo{author}{\bibfnamefont{L.}~\bibnamefont{Sangaletti}},
 \bibinfo{author}{\bibfnamefont{A.}~\bibnamefont{Goldoni}},
 \bibinfo{author}{\bibfnamefont{U.}~\bibnamefont{del Pennino}},
 \bibinfo{author}{\bibfnamefont{C.}~\bibnamefont{Kim}},
 \bibinfo{author}{\bibfnamefont{Z.-X.}~\bibnamefont{Shen}},
 \bibinfo{author}{\bibfnamefont{A.}~\bibnamefont{Revcolevschi}},
  \bibnamefont{and}
  \bibinfo{author}{\bibfnamefont{G.}~\bibnamefont{Dhalenne}},
  \bibinfo{journal}{Phys. Rev. B} \textbf{\bibinfo{volume}{55}},
  \bibinfo{pages}{1459} (\bibinfo{year}{1996}).


\bibitem{footnoteerror} The error bars take into account the differences 
between the two Li(1) models (see Fig. \ref{li-models}, top) and the 
numerical errors of the fitting procedure.



\bibitem{footnote7}We doubled the unit cell along
the chain direction and stabilized the following periodic spin
arrangements to calculate $J_1$ and $J_2$:
$\uparrow\uparrow\uparrow\uparrow,
\uparrow\downarrow\uparrow\downarrow,
\downarrow\uparrow\uparrow\uparrow$.

\bibitem{footnoteSLD} For $U_d= 5.5$\,eV and $U_d= 8.0$\,eV
we obtain $J_1=-16.7$\,meV, $J_2=4.1$\,meV, $\alpha=-0.25$ and
  $J_1=-10.1$\,meV, $J_2=1.9$\,meV, $\alpha=-0.19$, respectively.


\bibitem[{\citenamefont{Kanamori}(2008)}]{Tarui08}
\bibinfo{author}{\bibfnamefont{J.}~\bibnamefont{Kanamori}},
  \bibinfo{journal}{J. Phys. Soc. Jpn.} \textbf{\bibinfo{volume}{77}},
  \bibinfo{pages}{043703} (\bibinfo{year}{2008}).

\bibitem[{\citenamefont{Vavilova et~al.}(2008)\citenamefont{Vavilova, Moskvin,
  Arango, Sotnikov, Kataev, Drechsler, Volkova, Vasiliev, and
  B{\"u}chner}}]{Vavilova08}
\bibinfo{author}{\bibfnamefont{E.}~\bibnamefont{Vavilova}},
  \bibinfo{author}{\bibfnamefont{A.}~\bibnamefont{Moskvin}},
  \bibinfo{author}{\bibfnamefont{Y.}~\bibnamefont{Arango}},
  \bibinfo{author}{\bibfnamefont{A.}~\bibnamefont{Sotnikov}},
  \bibinfo{author}{\bibfnamefont{V.}~\bibnamefont{Kataev}},
  \bibinfo{author}{\bibfnamefont{S.-L.} \bibnamefont{Drechsler}},
  \bibinfo{author}{\bibfnamefont{O.}~\bibnamefont{Volkova}},
  \bibinfo{author}{\bibfnamefont{A.}~\bibnamefont{Vasiliev}}, \bibnamefont{and}
  \bibinfo{author}{\bibfnamefont{B.}~\bibnamefont{B{\"u}chner}},
  \bibinfo{journal}{arXiv:} p. \bibinfo{pages}{0810.5754v2}
  (\bibinfo{year}{2008}).

\bibitem{footnote8}Besides the very small energy
differences compared to the large total energy for different Li(1)
elongations from the equilibrium position, the main problem for larger
cells is the rather bad convergency behavior. The bad convergency is
related to the well known problem of ``charge shuffling'' between
(with respect to their charge density) almost identical, but
crystallographically formaly different atoms.


\bibitem[{\citenamefont{M\'alek et~al.}(2008)\citenamefont{M\'alek, Drechsler,
  Nitzsche, Rosner, and Eschrig}}]{Malek08}
\bibinfo{author}{\bibfnamefont{J.}~\bibnamefont{M\'alek}},
  \bibinfo{author}{\bibfnamefont{S.-L.} \bibnamefont{Drechsler}},
  \bibinfo{author}{\bibfnamefont{U.}~\bibnamefont{Nitzsche}},
  \bibinfo{author}{\bibfnamefont{H.}~\bibnamefont{Rosner}}, \bibnamefont{and}
  \bibinfo{author}{\bibfnamefont{H.}~\bibnamefont{Eschrig}},
  \bibinfo{journal}{Phys.\ Rev. B} \textbf{\bibinfo{volume}{78}},
  \bibinfo{pages}{060508(R)} (\bibinfo{year}{2008}).

\bibitem[{\citenamefont{M\'alek et~al.}(2009)\citenamefont{M\'alek, Kuzian,
  Nishimoto, Knupfer, and Drechsler}}]{Malek1}
\bibinfo{author}{\bibfnamefont{J.}~\bibnamefont{M\'alek}},
  \bibinfo{author}{\bibfnamefont{R.}~\bibnamefont{Kuzian}},
  \bibinfo{author}{\bibfnamefont{S.}~\bibnamefont{Nishimoto}},
  \bibinfo{author}{\bibfnamefont{M.}~\bibnamefont{Knupfer}}, \bibnamefont{and}
  \bibinfo{author}{\bibfnamefont{S.-L.} \bibnamefont{Drechsler}},
  \bibinfo{journal}{(in preparation)}  (\bibinfo{year}{2009}).

\bibitem[{rem()}]{remark}
\bibinfo{note}{The remaining parameters of the five-band extended Hubbard model
  read $U_d$=8.5~eV, $U_p=4.1$~eV, $U_{pp}=2.9$~eV, $K_p=0.6$~eV 
(Hund's rule coupling on O-sites), $V_{pd}=0.65$~eV (inter-site Coulomb 
interaction, neglected in Ref.~\onlinecite{Malek08}), $\varepsilon_{d}=0$, 
   $\varepsilon_{p_x}=3.6$~eV (in-chain) and $\varepsilon_{p_y}=3.4$~eV 
  (perpendicular to the chain). The transfer integrals read
  $t_{p_yd}=0.662$~eV, $t_{p_xd}=0.765$~eV, $t_{p_xp_x}=0.84$~eV, $t_{p_yp_y}=-
  t_{p_xp_x}/4$ in chain direction as well as 0.96~eV and -0.24~eV in
  transversal direction (see also Ref.~\onlinecite{Mizuno99},
  notice the different notation of the chain axis (see Fig.\ 6).}

\bibitem{footnote9}The accuracy of the least-square fit for the 
differences between the Hubbard and Heisenberg
eigen energies amounts 10$^{-6}$.


\bibitem[{\citenamefont{Mizuno et~al.}(1998)\citenamefont{Mizuno, Tohyama,
  Maekawa, Osafune, Motoyama, Eisaki, and Uchida}}]{Mizuno99}
\bibinfo{author}{\bibfnamefont{Y.}~\bibnamefont{Mizuno}},
  \bibinfo{author}{\bibfnamefont{T.}~\bibnamefont{Tohyama}},
  \bibinfo{author}{\bibfnamefont{S.}~\bibnamefont{Maekawa}},
  \bibinfo{author}{\bibfnamefont{T.}~\bibnamefont{Osafune}},
  \bibinfo{author}{\bibfnamefont{N.} \bibnamefont{Motoyama}},
  \bibinfo{author}{\bibfnamefont{H.}~\bibnamefont{Eisaki}}, \bibnamefont{and}
  \bibinfo{author}{\bibfnamefont{S.}~\bibnamefont{Uchida}},
  \bibinfo{journal}{Phys.\ Rev. B} \textbf{\bibinfo{volume}{57}},
  \bibinfo{pages}{5326} (\bibinfo{year}{1998}).

\bibitem[{\citenamefont{Eskes and Jefferson}(1993)}]{Eskes93}
\bibinfo{author}{\bibfnamefont{E.}~\bibnamefont{Eskes}} \bibnamefont{and}
  \bibinfo{author}{\bibfnamefont{A.}~\bibnamefont{Jefferson}},
  \bibinfo{journal}{Phys.\ Rev.\ B} \textbf{\bibinfo{volume}{48}},
  \bibinfo{pages}{9788} (\bibinfo{year}{1993}).

\bibitem[{\citenamefont{Braden et~al.}(1996)\citenamefont{Braden, Wilkendorf,
  Lorenzana, Ain, McIntyre, Behruzi, Heger, Dhalenne, and
  Revolevschi}}]{Braden96}
\bibinfo{author}{\bibfnamefont{M.}~\bibnamefont{Braden}},
  \bibinfo{author}{\bibfnamefont{G.}~\bibnamefont{Wilkendorf}},
  \bibinfo{author}{\bibfnamefont{J.}~\bibnamefont{Lorenzana}},
  \bibinfo{author}{\bibfnamefont{M.}~\bibnamefont{Ain}},
  \bibinfo{author}{\bibfnamefont{G.}~\bibnamefont{McIntyre}},
  \bibinfo{author}{\bibfnamefont{M.}~\bibnamefont{Behruzi}},
  \bibinfo{author}{\bibfnamefont{G.}~\bibnamefont{Heger}},
  \bibinfo{author}{\bibfnamefont{G.}~\bibnamefont{Dhalenne}}, \bibnamefont{and}
  \bibinfo{author}{\bibfnamefont{A.}~\bibnamefont{Revolevschi}},
  \bibinfo{journal}{Phys.\ Rev.\ B} \textbf{\bibinfo{volume}{54}},
  \bibinfo{pages}{1105} (\bibinfo{year}{1996}).

\bibitem{footnote10}Similar problems occur also for $J_2$ except the
fact that it is hardly affected by $K_{pd}$ but by the much smaller
value $K_{pp}$.  In the present calculation we have therefore ignored
$K_{pp}$ as it is usually done in the literature.


\bibitem[{\citenamefont{Weisse et~al.}(1999)\citenamefont{Weisse, Wellein, and
  Fehske}}]{Weisse99}
\bibinfo{author}{\bibfnamefont{A.}~\bibnamefont{Weisse}},
  \bibinfo{author}{\bibfnamefont{G.}~\bibnamefont{Wellein}}, \bibnamefont{and}
  \bibinfo{author}{\bibfnamefont{H.}~\bibnamefont{Fehske}},
  \bibinfo{journal}{Phys.\ Rev.\ B} \textbf{\bibinfo{volume}{60}},
  \bibinfo{pages}{6566} (\bibinfo{year}{1999}).

\end{thebibliography}
%
\end{document}